\documentclass[a4paper]{jpconf}
\usepackage{graphicx}

\begin{document}
\title{SQM2016: Theory Summary}

\author{Vincenzo Greco}

\address{Department of Physics and Astronomy, University of Catania, Via S. Sofia 64, I-95125 Catania, IT}
\ead{greco@lns.infn.it}

\begin{abstract}
This is an overview of the main theoretical developments presented at SQM2016 held in Berkeley, California (USA) in June 2015.
\end{abstract}

\section{Introduction}
During the \textit{ "Strangeness in Quark Matter 2016"} (SQM2016) conference, more than 50 talks were presented 
about the theoretical effort to understand the many facets of the heavy-ion collisions and the properties of the hot
QCD matter. Here in a few pages we overview the main aspects that have been subject of intense discussion focusing
on the theoretical plenary talks.

\section{Collectivity in Small Systems}
The study of several observables and their critical assessment impinges on the understanding of the initial state
space-momentum distribution and correlations. At this conference it was presented a main advancement
showing that strong eccentric initial fluctuations of the proton shape,  once coupled to viscous hydrodynamics,
can account for the $v_2$ observed at $p+Pb$ collision at LHC \cite{Mantysaari:2016ykx}. A main novelty is that proton shape fluctuations
can be constrained by exclusive vector meson production in lepton-proton scatterings at HERA lepton-proton scattering, creating
a nice link with high-energy hadronic physics. It remains the issue of the validity of hydrodynamics for a system
with large initial Kundsen number. Although the system may undergo "hydrodynamization" in a short time scale
as indicated by AdS/CFT approaches \cite{Schenke}, it would be important to see if kinetic theory also leads to a similar conclusions.
However still such an approach has to be contrasted with the relevance of local initial momentum correlations (on a $1/Q_s$ scale) 
that lead also to final anisotropies, but not directly related to the space eccentricities \cite{Lappi:2015vta}.

\section{Sub-threshold Strangeness and Charm production}
The large $\phi/K^-$ and $\Xi^-/\Lambda$ ratios reported by HADES and FOPI \cite{Agakishiev:2009ar,Agakishiev:2009rr} at subthreshold energy
have not been correctly predicted by microscopic transport model and the thermal fits are also not convincing.
It has been shown, employing uRQMD augmented by $N^* \rightarrow \phi, \Xi^-$ decays, that the Fermi motion plus secondary 
interaction are a fast way to generate collisions pairs with invariant mass that goes above the threshold. 
Such a mechanism appears to correctly
reproduce the low $\sqrt{s_{NN}}$ enhancement measured experimentally \cite{Steinheimer}, predicting a peak at
 $\sqrt{s_{NN}}=1.25 \,\rm GeV$ . The quantitative prediction is reasonably
regulated by the ANKE pp data for the $\phi$ production in the $2.5-3$ GeV energy range \cite{Hartmann:2012ia}.
A similar study for subthreshold charm production has also been presented, showing that 
$J/\Psi, \Lambda_c, \overline D$ may have the chance to be abundant already at SIS100
energy range, $E_{lab}=5-11 \, \rm AGeV$. Finally it is becoming clear that fast resonance excitation and decays
allow  for an effective thermalization of hadron yields in a time scale comparable with
the nuclei passing time, thanks to the large density of states \cite{Steinheimer:2016vzu}.

\section{Thermal model, freeze-out, susceptibilities and phase diagram}
The statistical thermal hadronization model (SHM) is able to correctly predict the hadronic yields, and surprisingly recently
also light nuclei like $^3He, ^3_\Lambda He$  and their anti-nuclei. Considering that this involves 
a span of about 9 orders of magnitude and is valid at least over the entire range of collisions from SPS
to LHC energy \cite{Andronic}, this result constitutes one of the bulk achievements of the field that easily at a glance
shows that a matter hadronizing with a temperature $T\simeq 155-165 \, MeV$ has been created \cite{Andronic:2008gu}. 
At first glance it always appears quite surprising that the hadronic system can achieve a nearly full
equilibrium in a short time. At this conference it has been shown that a microscopic mechanism like the Hagedorn state decays, at the
energy densities relevant for QCD transition, can lead to a system nearly born in chemical equilibrium \cite{Beitel:2016ghw}.

It is however known that there is a tension in the data on $p,\overline p, \Lambda$ production at LHC. This is certainly 
very interesting and according to some of the talks may signal hadronic chemical re-interaction \cite{Becattini:2014hla, Stock}
that shifts the apparent temperature. While this could be a quite natural explanation, an important warning
comes from the most recent lattice QCD results on susceptibilities, which has been an hot topic of the conference.
From lQCD study of the susceptibilities, involving the strange quantum number and in particular the 
Koch' ratio \cite{Karsch}, it appears clear that the Hadron Resonance Gas (HRG)  model, underlying the SHM, from the PDG
is not able to reproduce the Koch's ratio,$\chi_{11}^{BS}/\chi_2^S$, as evaluated from lQCD for $T\geq 160 \, \rm MeV$. 
A new result presented also shows that even a HRG based
on Quark Model that appears to be in a quite good agreement with several lQCD calculations, misses for
example the $\chi_4^S/\chi_2^S $ already at $T> 140 \,\rm MeV$, probably because of lack of $s=1$ meson states \cite{Ratti, Noronha}.

A fundamental part of the phase diagram is certainly the Equation of State (EoS) at finite $\mu_B/T$.
The Wuppertal-Budapest Collaboration has shown to be able to evaluate it up to $\mu_B/T < 2.5$
and up to term of the $(\mu_B/T)^6$ order \cite{Ratti}.

A key part of the relativistic HIC program is certainly the search for the QCD critical point. It has been shown that
kurtosis $\kappa\sigma^2$ of net protons measured by STAR and $\chi_4/\chi_2$ in lQCD appear to be in good agreement within the
present uncertainties in the region $\sqrt{s_{NN}}> 19 \,\rm GeV$ corresponding to $\mu_B/T<2$, where lQCD results
are reliable \cite{Karsch}. It has to be mentioned that in this region lQCD does not see a critical point.
Dynamical studies of critical fluctuations and of resonance decay impact have been presented, 
assuming a critical point. They show a robust remnant of critical fluctuations \cite{Bluhm}, but further advancements
are needed before a reliable comparison can be done.
Soon the entire range accessible with BES-I and II at RHIC will be accessible through Taylor expansion in lQCD.
A proposal to face the search for the critical point by looking the high energy collisions versus the rapidity
has been presented in a parallel session\cite{Kapusta}.

\section{Heavy Flavor}
The heavy flavor attracted quite some interest already more than a decade ago because of the large suppression
observed with a call for large non-perturbative effects at least for charm quarks with $p_T \leq 10\,\rm GeV$.
A main difference from the physics of the bulk QGP matter comes from the fact that the initial $p_T$ distribution 
can be evaluated in a pQCD scheme and in particular the FONLL has been show to be in agreement with the
experimental data within the inherent uncertainties. Furthermore one expects marginal thermal production,
which is confirmed, within current error bars, also by the observed $N_{coll}$ scaling.
A main challenge, currently, is the simultaneous description of both the $R_{AA}(p_T)$ and the $v_2(p_T)$.
It has been pointed out that several ingredients contribute to this, but the main impact can come from
the $T$ dependence of the transport coefficient. In particular a constant drag coefficient $\gamma $ leads
to a more efficient building-up of $v_2$ especially in the low-intermediate $p_T \leq 4 \, \rm GeV$.
There a weak dependence of the drag as predicted by Quasi-Particle models \cite{Berrehrah:2013mua,Das:2015ana} 
or by T-matrix approach \cite{vanHees:2007me}
leads to $v_2$ about a factor 2-3 larger w.r.t. the $T^2$ dependence as in pQCD ($\alpha_s=cost.$) or
AdS/CFT. A similar trend is observed also for $v_3$ for heavy quarks at intermediate-high $p_T$, as discussed in \cite{Prado}.

Another source of uncertainty comes from the microscopic heavy quark dynamics.
A real heavy quark would undergo a Brownian motion  well described by Langevin transport equations.
A charm quark can be envisaged to be marginally heavy, if one considers that while for
the thermal production  $m_c$ has to be compared with $T$, for the scatterings
other scales enters the problem, $gT$ and/or $\pi\,T$, that are comparable to
the charm mass at temperatures $T\sim 300-400 \, \rm MeV$,
reached in particular at LHC energy.
A direct comparison between Langevin and Boltzmann dynamics shows non sizeable differences
for bottom quarks, while for charm it can lead to differences that range between $10-30\%$
in the intermediate $p_T$ region for spectra and elliptic flow, depending on the specific model for the
in-medium scatterings considered \cite{Das:2013kea}.
The effect is predicted to be larger for $c-\overline c$ angular correlations.
In the low $p_T$ limit the differences between the two approaches become marginal
for the predictions on the charm spectra (or $R_{AA}$) that especially at LHC energy appear
to reach thermalization when a thermal model is able to correctly describe $R_{AA}(p_T)$ for $p_T < 2 \, \rm GeV$.

The comparison to data favours a space-diffusion coefficient $2\pi TD_s$ that rises with temperature $T$ 
in agreement with lQCD calculation, although such a statement suffers from the systematic errors in both lQCD 
and the phenomenological models. In the latter the main sources
of difference may come from the background bulk expansion and from details in the hadronization
mechanism. However, in general, it appears that models including quark coalescence are able to provide predictions closer
to the experimental data at both RHIC and LHC energy.

An important advancement presented has been the Linearized Boltzmann Transport (LBT), from the 
LBL-CCNU Collaboration, that is able to treat
in the same framework both the light and heavy hadrons suppression \cite{Cao:2016gvr}, confirming in the heavy quark sector the
necessity to have a modification of the the diffusion coefficient w.r.t. the $T^2$ dependence, as pointed out in \cite{Das:2015ana}.
It is interesting to notice that a similar dependence for the diffusions coefficient appears to be necessary
also for the $R_{AA}-v_2$ for high $p_T$ minijets as pointed out by the CUJET approach \cite{Xu:2015bbz}.
In the latter the microscopic mechanism of a minimum in $ 2 \pi T D_s$ would be driven by the presence
of monopoles in the bulk medium, while for heavy quarks in the low $p_T$ region it would be consistent 
with the dynamics implied by the remnants of confinement according to TAMU T-matrix approach \cite{Liu:2016nwj,Liu}.

A novel aspect of the heavy quarks physics that has been started to be investigated is the impact of
the strong initial magnetic field on their dynamical evolution.
Some first work shows a modification of the parallel and transverse HQ diffusion coefficient that becomes
significant for value of the magnetic field $eB> T^2 $ \cite{Yee,Fukushima:2015wck}, which however may marginally occur in HIC at current energy.

On the other hand it has been shown initial $B$ entails a sizeable directed flow ($v_1$) of charm quarks (CQs) 
much larger than light quarks due to a combination of several favorable conditions for CQs, mainly: (i) unlike light quarks 
formation time scale of CQs, $\tau_f \simeq \, 0.1 \rm fm/c$ is comparable to the time scale when
$ B$ is around its maximum value and  (ii) the kinetic relaxation time of CQs  
is similar to  the QGP lifetime \cite{Das:2016cwd}. The effect is also odd under charge exchange allowing to distinguish it from the vorticity 
of the bulk matter due to the initial angular momentum conservation. 

\section{Quarkonia}
The medium modification of the quarkonia production is certainly a key probe of the deconfined matter.
During the years and going to collisions at increasing energies it has become clear that for charmonia
more than the suppression it is the regeneration that plays a key role for the understanding of its
production especially at top LHC energy. The large amount of data collected as a function of rapidity and centralities
clearly show this. The theoretical approaches based on a transport evolution including both
regeneration and dissociation are able to predict the $J/\Psi$ production fairly well \cite{Zhou,Liu:2009nb,Zhao:2011cv}. 
However the impact of the uncertainties  in the charm cross sectio still leaves open a key question:
is charmonia yield
consistent with a regeneration at the chemical freeze-out temperature as for light hadron production? or
it is necessary to have a dynamical description of suppression and regeneration to understand the production in the QGP matter?
From the point of view of the SHM a clear prediction would be a strong reduction of the ratio $\Psi(2S)/J/\Psi$  with
respect to pp and pA \cite{Andronic:2009sv}. From the point of view of the dynamical regeneration approach a signature of the dynamical
nature of the production is the anomalous regeneration of $\Psi(2S)$ observed in central $Pb+Pb$ collisions
at high $p_T>3 \,\rm GeV$, a behavior that is predicted in such a picture due to the delayed recombination into $\Psi(2S)$ 
w.r.t. $J/\Psi$ and a larger inherited radial flow \cite{Du}. 
This implies a quite different behavior of $R_{AA}(p_T)$ for the  $J/\Psi$ w.r.t. $\Psi(2S)$ with the latter showing a peak
at $p_T \simeq 4\,\rm GeV$.
Upcoming experimental data at low $p_T$ will allow to clarify
this issue. Certainly it remains important to measure the $v_2$ of the $J/\Psi$ with significantly
reduced statical and systematic errors. This would allow, independently
of the measurement of the charm cross section, the dynamics of the $J/\Psi$ production to be clarified, 
as anticipated in \cite{Greco:2003vf}.

An important theoretical development toward a quantum treatment of the quarkonia suppression has been presented.
In fact most of the approaches treat the several quarkonia states as independent states by means
of decays width, while they are different excited quantum states of the same system.
In the framework of a Schroedinger-Langevin equation that encodes decoherence dynamics \cite{Gossiaux:2016htk}.
A first comparison show a reasonable agreement with data on bottomonia suppression except for most central collisions.
However before full realistic calculation, the impact of realistic 3D potential 
extracted from lQCD and the impact of the assumptions for
the initial quantum states of quarkonia as well as the relation to lQCD spectral functions, still have to be investigated \cite{Gossiaux}.

\section{Chiral Anomaly}
The Chiral Magnetic Effect (CME) is a remarkable phenomenon emerging from a highly nontrivial interplay of QCD
chiral symmetry, axial anomaly, and gluonic topology. 
The heavy ion collisions supply a hot chiral-symmetric QGP under a strong magnetic field,
hence gluonic topological fluctuations could generate a chirality imbalance observable experimentally
as a charge separation along the direction of the magnetic field.
Experimentally a signal of such a separation has been measured especially in the energy range of 
BES-RHIC at $\sqrt{s_{NN}}=7.7-62.4\,\rm GeV$ \cite{Kharzeev:2015znc}, although currently it seems that
models with magnetic field-independent flow backgrounds can also be
tuned to reproduce the current measurements.

Indeed on the theoretical side a transport theory that includes
the dynamics of chiral currents and the associated anomaly, is needed. This is leading to the
development of a chiral viscous hydrodynamics and some attempts to develop a chiral kinetic transport
theory; the latter would allow studying the pre-thermal glasma stage that is likely to be the most
pertinent one given the short life-time of the magnetic field. Such efforts will supply the necessary
theoretical tools to predict the size and the evolution of the CME with energy and centrality.
Also it can allow us to study in a self-consistent way several observables related to CME and to see
if from the experimental data  a coherent picture for a fundamental effect emerges, which however till now
is associated to a weak signal over a large background. Nicely,
a first attempt to describe the centrality dependence of the charge separation within chiral hydro
appears to be consistent with the dynamics implied by the CME \cite{Liao}.

\section{Global Polarization}
An aspect of relativistic HIC that has been mostly overlooked is that in non-central collisions
due to the large orbital momentum one is likely to create a system with a huge vorticity.
One can envisage that particles emerging from such
a highly vorticous fluid are globally polarized with their spins on average pointing
along the system angular momentum.
To study such a dynamics the development of a self-consistent theory that treat 
vorticity in a relativistic framework is needed. This has been recently investigated in the context of
the QGP physics assuming a thermal fluid deriving a general formula that relates the vorticity
of a fluid to its polarization for particles of spin $1/2$ \cite{Becattini:2013fla}. This has allowed embedding the polarization
dynamics into the ECHO-QGP viscous hydro code for RHIC's studying in particular the $\Lambda$ polarization
that is accessible experimentally \cite{Becattini:2015ska,Karpenko}. 
Distinctive feature of thermodynamic polarization would be a  C-even
effect, that particle and antiparticle have the same polarization, unlike e.m. induced polarization.
The predictions appear to follow a similar increase of the polarization with decreasing beam energy.
Interestingly in experiments the polarization of the order of a few percent appears to be even larger
than the prediction \cite{Karpenko}. However the effect of resonance decays, hadronic rescattering, impact of the Pauli Blocking \cite{Huang}
have to be investigated. Certainly the developments of such studies has opened a new direction
and can allow us to have an insight on the vorticity of the matter created in the early stage of the collision,
adding a new dimension in the HIC phase space.

\section{Summary}
In Summary,
there is a clear progress in the understanding of the QCD matter at high temperature created in relativistic heavy-ion
collisions. Certainly in the last decade an understanding of the main features of the collision dynamics and
its collective expansion has been achieved, and the surprising behavior of several observables in pA from strangeness
production to anisotropic flows will at the end allow us
to have deeper understanding of the matter created in such collisions and the properties of hot QCD.
At the some time one should appreciate that
many questions were not even conceivable a decade ago, but are now under active investigation thanks to 
the progress that the field has achieved in the understanding of the bulk properties of the matter created.

\section{Acknowledgments}
V.G. is supported under QGPDyn ERC Grant no. 259684.

\end{document}